\documentclass[prx,twocolumn,floatfix,superscriptaddress,longbibliography]{revtex4-1}

\usepackage{amssymb,amsmath,amstext}
\usepackage{graphicx}
\usepackage{epstopdf}
\usepackage{color}
\usepackage{bm}
\usepackage{appendix}
\usepackage[T1]{fontenc}
\usepackage{bbold}
\usepackage{bbm}
\usepackage{latexsym}
\usepackage{placeins}
\usepackage{enumitem}
\usepackage[colorlinks=true,citecolor=blue,linkcolor=magenta]{hyperref}
\usepackage[dvipsnames]{xcolor}
\usepackage{placeins}

\renewcommand{\vec}[1]{\boldsymbol{#1}} 
\newcommand{\ket}[1]{|#1\rangle} 

\newcommand{\mi}{\mathrm{i}}

\begin{document}

\title{Efficient Online Quantum Circuit Learning with No Upfront Training}

\author{Tom O'Leary}
\thanks{The first two authors contributed equally to this work.}
\affiliation{Theoretical Division, Los Alamos National Laboratory, Los Alamos, NM, USA.}
\affiliation{Department of Physics, Clarendon Laboratory, University of Oxford, Oxford, UK.}

\author{Piotr Czarnik} 
\thanks{The first two authors contributed equally to this work.}
\affiliation{Institute of Theoretical Physics, Jagiellonian University, Krak\'ow, Poland.}
\affiliation{Mark Kac Center for Complex Systems Research, Jagiellonian University, Krak\'ow, Poland}

\author{Elijah Pelofske}
\affiliation{Information Systems \& Modeling, Los Alamos National Laboratory, Los Alamos, NM, USA.}

\author{Andrew T. Sornborger} 
\affiliation{Information Sciences, Los Alamos National Laboratory, Los Alamos, NM, USA.}
\affiliation{Quantum Science Center, Oak Ridge, TN 37931, USA.}

\author{Michael McKerns} 
\affiliation{Information Sciences, Los Alamos National Laboratory, Los Alamos, NM, USA.}
\affiliation{The Uncertainty Quantification Foundation, Wilmington, DE 19801, USA.}

\author{Lukasz Cincio}
\affiliation{Theoretical Division, Los Alamos National Laboratory, Los Alamos, NM, USA.}
\affiliation{Quantum Science Center, Oak Ridge, TN 37931, USA.}

\begin{abstract}
We propose a surrogate-based method for optimizing parameterized quantum circuits which is designed to operate with few calls to a quantum computer.
We employ a computationally inexpensive classical surrogate to approximate the cost function of a variational quantum algorithm.
An initial surrogate is fit to data obtained by sparse sampling of the true cost function using noisy quantum computers. 
The surrogate is iteratively refined by querying the true cost at the surrogate optima, then using radial basis function interpolation with existing and new true cost data.
The use of radial basis function interpolation enables surrogate construction without hyperparameters to pre-train. 
Additionally, using the surrogate as an acquisition function focuses hardware queries in the vicinity of the true optima.
For 16-qubit random 3-regular Max-Cut problems solved using the QAOA ansatz, we find that our method outperforms the prior state of the art. 
Furthermore, we demonstrate successful 
optimization of QAOA circuits for 127-
qubit random Ising models on an IBM 
quantum processor using measurement counts of the order of $10^4-10^5$.
The strong empirical performance of this approach is an important step towards the large-scale practical application of variational quantum algorithms and a clear demonstration of the effectiveness of classical-surrogate-based learning approaches.

\end{abstract}
\maketitle

\section{Introduction}
\label{section:introduction}

The prospect of a quantum computer providing a computational advantage over a classical computer
is of broad interest~\cite{farhi2014quantum,huang2020predicting,caro2021generalization,childs2018toward}. 
Quantum processors have developed to the point of approaching the limit of classical simulability ~\cite{arute2019quantum, kim2023evidence, aharonov2023polynomial, rudolph2023classical, tindall2023efficient, 
schuster2024polynomial,
beguvsic2024fast,anschuetz2022efficient, cerezo2023does}.
However, decoherence, cross-talk error, imperfect gate calibration, and measurement error, collectively referred to as noise, place limitations on current quantum devices. 
Variational quantum algorithms (VQAs) are a widely studied approach to utilizing pre-fault-tolerant noisy quantum computers~\cite{cerezo2020variationalreview}, which we expect to be of use in the fault-tolerant era.
VQAs are generally formulated as optimization problems, with problem solutions encoded into the minima of a cost function. 
Parameterized quantum circuits (PQCs) are iteratively used to prepare candidate solutions until a cost function minimum (ideally a global minimum) is found. 
A~classical computer updates the quantum circuit parameters at each iteration based on previously obtained cost function values. 
In order to minimize the effects of noise, VQAs use low-depth quantum circuits, typically scaling at most logarithmically in system size for near-term quantum computers.

\begin{figure}[t]
        \includegraphics[width=\columnwidth]{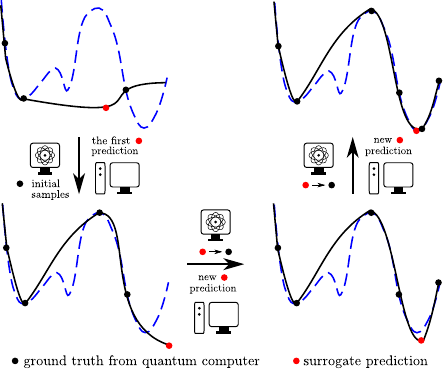}
    \caption{{\bf Surrogate-based optimization of parametrized quantum circuits}. We use an inexpensive classical surrogate $C^{\rm surr}$ (black curve) to approximate the true cost function $C$ (dashed blue curve) that is sampled using a quantum computer (black dots). Top-left: we fit the first surrogate using randomly sampled points from the quantum computer.    
    We then find the surrogate minimum to determine the next point to sample on the quantum hardware (red dot). Bottom-left: we use the new point to refine our surrogate. Right column: this procedure is iterated until a good approximation of the true cost minimum is found.} 
    \label{fig:method}
\end{figure}

There are several challenges to obtaining a quantum advantage with this approach. 
In particular, for a broad range of expressive PQCs, or PQCs affected by unital noise, one encounters cost function landscapes which concentrate to their average value exponentially fast in the problem size~\cite{mcclean2018barren, wang2020noise, cerezo2020cost, arrasmith2021equivalence, holmes2021connecting,martin2022barren}.  
These cost function barren plateaus require the cost function to be sampled a prohibitive number of times to be optimized. 
Even noise-free cost functions with non-vanishing gradients present a challenging optimization problem, due to the rapidly increasing number of local minima with PQC depth~\cite{bittel2021training, anschuetz2022beyond}.
Nevertheless, PQCs with a provable absence of barren plateaus and special optimization initialization strategies have been proposed to mitigate these issues~\cite{pesah2020absence, sack2022avoiding, schatzki2022theoretical, zhang2023absence,zhang2022escaping,park2023hamiltonian,wang2023trainability,park2024hardware,shi2024avoiding,puig2024variational,cao2024exploiting,mele2022avoiding}. 
Furthermore, it has recently been shown that cost function landscape regions without barren plateaus can be efficiently approximated classically, if access to quantum cost function data in these regions is available~\cite{cerezo2023does}.
Moreover, it has been shown that such data can be obtained efficiently, with a shot cost scaling at most polynomially with the number of PQC parameters~\cite{lerch2024efficient,bermejo2024quantum}. In other words, a classical surrogate of the quantum cost function can be constructed efficiently. Motivated by these developments, we propose a surrogate-based approach, which seeks to leverage recent progress \cite{diaw2024efficient} in surrogate-based optimization methods to enable on-device optimization of PQCs for challenging problems.

Surrogate methods
attempt to minimize the number of evaluations of a computationally expensive cost function, $C$, by constructing an inexpensive surrogate cost function, $C^{\rm surr}$,
to serve as a lightweight 
approximation of $C$.
If evaluations of the surrogate are computationally negligible, then the computational cost of optimization strongly depends on how efficiently the surrogate method can produce an accurate $C^{\rm surr}$ with a minimal number of calls to $C$. 
When $C^{\rm surr}$ has a form that can approximate $C$ sufficiently well, surrogate-based optimizations of $C$ can require orders of magnitude fewer evaluations of the true cost function, while yielding an accurate solution~\cite{mockus1975bayes, mockus1991bayesian}.

Here, we propose a new surrogate approach for the optimization of quantum circuits. We will use a computationally inexpensive classical surrogate to approximate a VQA cost function. We seek to minimize the total number of shots required to optimize a PQC.
We use an approach that attempts to efficiently construct a surrogate cost function on-the-fly, without requiring any pre-training of the surrogate. Such an approach is of particular importance for tackling problems that can not be simulated classically on 
physical devices with time-variable noise. 
In such a case, no prior knowledge about the device cost function landscape may be available. 

There is a growing body of work using classical surrogate models in VQAs
~\cite{otterbach2017unsupervised,sung2020using,shaffer2023surrogate,muller2022accelerating,cheng2024quantum,kim2023quantum}. Most of these approaches
use Bayesian Optimization~\cite{mockus1975bayes, mockus1991bayesian} with Gaussian Process models~\cite{rasmussen2004gaussian} as surrogates. The surrogate parameters are typically selected after performing upfront training, then the pre-trained surrogate is used to determine points to be evaluated on quantum hardware ~\cite{otterbach2017unsupervised,shaffer2023surrogate,muller2022accelerating,cheng2024quantum,kim2023quantum}.
Some previous work also uses trust regions to constrain the parameter space for which the surrogate is valid~\cite{sung2020using,shaffer2023surrogate,cheng2024quantum}. 
Another approach is to use a surrogate obtained exclusively through an approximate numerical simulation of $C$~\cite{gustafson2024surrogate,lerch2024efficient}, however, without using quantum data its generality is limited~\cite{lerch2024efficient}.

From an optimization theory perspective, most modern surrogate-based approaches use a sampling method that relies on uncertainty reduction to select the new training points used to refine the surrogate. 
Both active learning \cite{cohn1994improving} and Bayesian Optimization \cite{mockus1975bayes, mockus1991bayesian} commonly choose Gaussian Process \cite{rasmussen2004gaussian} models as surrogates, as a Gaussian Process model can generate a mean and a variance, therefore yielding an uncertainty estimate. 
Information about the true response function (from the training data) is iteratively folded into the surrogate through Bayes's rule, where the surrogate can be represented by the mean of the distribution in the Gaussian Process model, and is updated each iteration when the posterior is used to inform the prior for the next iteration.
Active learning and Bayesian Optimization perform optimization on an acquisition function within each iteration, to identify new training points which are used to inform the prior in the next iteration.
Surrogate-based approaches generally use bound-constrained local optimization algorithms, and lean heavily on incorporating information about the true response surface by iteratively informing the prior. 

While our approach also uses adaptive sampling to iteratively refine a surrogate, there are 
fundamental differences. 
We have previously shown that by applying 
known information about the problem through a scientific 
coordinate transform (essentially an operator transforming coordinate space to a constrained coordinate space) machine learning produces a surrogate whose form 
better approximates truth (\textit{i.e.} ground-truth on experimental data), and thus requires less training 
data to produce a surrogate that fits the training data well 
\cite{czarnik2022improving, mckerns2019optimal, mckerns2018automated}. Our method is built to leverage these 
transforms in a constrained pseudo-global optimization of the 
surrogate. 
More specifically, our acquisition function 
directly seeks $N$ extrema of the surrogate during each 
iteration, as opposed to indirectly through surrogate uncertainty reduction.
We then sample the corresponding points on truth, 
and use the error in prediction between the candidate 
extrema from the surrogate and truth at each iteration to 
better drive the surrogate towards truth, as shown schematically in Fig.~\ref{fig:method}. 
Notably, most implementations of Gaussian Process model-based approaches rely on uncertainty-based acquisition functions, 
making the implementation of such coordinate transforms challenging ~\cite{rasmussen2004gaussian}.
Unlike Gaussian Process model-based surrogates, our approach does not require hyperparameters specifying the correlations to the existing data points sampled from truth.
This enables more direct surrogate construction methods, such as radial basis function interpolation that also do not require pre-training to produce good quality surrogates.

We target the Quantum Approximate Optimization Algorithm (QAOA)~\cite{farhi2014quantum} as an example use-case of our surrogate-based optimization protocol. We apply QAOA to the Max-Cut combinatorial optimization problem~\cite{crooks2018performance,boulebnane2021predicting,wurtz2021maxcut,sack2023large,santar2024squeezing,maciejewski2024multilevel} and finding low energy states of random-coefficient classical Ising models. 
Max-Cut problem instances are encoded as binary discrete polynomials known as quadratic unconstrained binary optimizations (QUBOs). The Max-Cut costs, or equivalently Ising model energies, are optimized with respect to parameters of the QAOA PQCs to find low-energy solutions.

For the Max-Cut problems we numerically simulate surrogate-based optimization for 16-qubit problem instances using measurement shot numbers typically available with cloud-based quantum computers. 
For this setup, we find that our approach outperforms a state-of-the-art VQA method (DARBO)~\cite{cheng2024quantum}. Next, we use numerical tensor network methods to simulate surrogate-based optimization of a shallow QAOA PQC for 127-qubit instances of a random Ising model on the heavy hex graph used by current IBM current quantum processors. We demonstrate a consistent improvement in solution quality over solutions obtained using parameter transfer from smaller problem instances found in previous work. 

We build on this by using the 
\texttt{ibm\_torino} quantum computer to apply our method to solve a 127-qubit Ising model instance. Again, we systematically improve upon previous parameter transfer results. 
To the best of our knowledge, this implementation is the largest on-device QAOA optimization in the literature, showcasing the potential of the surrogate-based approach. This performance gain is made possible by an extensive usage of $C^{\rm surr}$, as we evaluate it three orders of magnitude more frequently than the on-device cost, $C$.

We present our method in detail in Sec.~\ref{sec:alg}. In Sec.~\ref{sec:QAOA} we summarize the QAOA approach.  In Sec.~\ref{sec:numerical_results} we present the numerical results, and in Sec.~\ref{sec:dev_opt} we describe the hardware implementation of our approach. We conclude in Sec.~\ref{sec:conclusions}. 


\section{Our method}
\label{sec:alg}

Our goal is to optimize $C(\vec{\theta})$ with respect to a set of quantum circuit parameters, $\vec{\theta}=(\theta_1, \theta_2, \dots, \theta_N)$, keeping the number of $C(\vec{\theta})$ evaluations as small as possible. We assume that the parameters are bounded. We start with
an initial coarse sampling of the full parameter space to obtain $N_{\rm init}$ evaluations of $C(\vec{\theta})$. In other words, we evaluate the cost function for an initial set of parameters $\Theta_{\rm init}=\{\vec{\theta}_{1}, \vec{\theta}_2, \dots,  \vec{\theta}_{N_{\rm init}}\}$.  
If a heuristic choice of $\vec{\theta}$ is available, in particular when $C(\vec{\theta})$ is known to be a good parameter choice, one may include these parameters in $\Theta_{\rm init}$ to speed up optimization convergence. 
Otherwise, all initial parameters are obtained by random sampling. In Sec.~\ref{sec:MaxCut}, we use uniform random sampling of the parameter range to generate $\Theta_{\rm init}$, and supplement these random choices with heuristic parameters in Secs.~\ref{sec:QAOA_MPS} and~\ref{sec:dev_opt}.  

Next, we iterate the following steps until a termination criterion is met. 
\begin{enumerate}
\item 
We build a classical surrogate cost function $C^{\rm surr}(\vec{\theta})$ performing radial basis function interpolation using all $C(\vec{\theta})$ evaluations performed so far. More explicitly, we interpolate a set of pairs
\begin{displaymath}
\{ (\vec{\theta}, C(\vec{\theta})) \,|\, \vec{\theta} \in \Theta \},
\end{displaymath}
with $\Theta =\Theta_{\rm init}$ for the first iteration ($i=1$), and updated later as described below. The interpolation is performed with the \textit{mystic} scientific machine learning package~\cite{mckerns2011building,mckerns2009mystic}. 
\item 
We identify $N_{\rm opt}$ randomly chosen new points in the parameter space,
and use these points as starting values for
local minimizations (or maximizations) of $C^{\rm surr}(\vec{\theta})$. 
We identify the best of the optimization solutions as a candidate optimum, $\vec{\theta}_{\rm cand}^{(i)}$. 
\item
We evaluate $C(\vec{\theta}_{\rm cand}^{(i)})$, and check the termination condition. After this step $\Theta\leftarrow\Theta \cup \{\vec{\theta}_{\rm cand}^{(i)}\}$,  and $i\leftarrow i+1$.
\end{enumerate}
In this work, to demonstrate the method in a simple setup, we terminate the loop
after $N_{\rm it}$ iterations. We defer investigation of the more sophisticated termination conditions commonly used in classical surrogate approaches~\cite{diaw2024efficient} to future work. Our approach is depicted schematically in Fig.~\ref{fig:method}.

Our procedure outputs the best cost function value found on the noisy quantum computer, $C_{\rm opt}$, and the corresponding parameters, $\vec{\theta}_{\rm opt}$, where
\begin{align}
\vec{\theta}_{\rm opt} &= {\rm argmin}\, ({\rm argmax}) \, \{C(\vec{\theta}) \,|\, \theta \in  \Theta \},  \\ C_{\rm opt} &= C(\vec{\theta}_{\rm opt}),
\label{eq:Copt}
\end{align}
for minimization (maximization)
. Furthermore, in Secs.~\ref{sec:numerical_results} and~\ref{sec:dev_opt}, to improve the method's performance, we repeat the optimization runs multiple times, starting from different $\Theta_{\rm init}$ each time. 
Then, the outcomes of the best run are accepted as the optimization output. Such a strategy is commonly used by state-of-the-art optimization algorithms in the case of cost function landscapes with multiple local optima~\cite{cheng2024quantum}. 


\section{QAOA}
\label{sec:QAOA}
In  QAOA, we start with a~classical Hamiltonian that typically describes a combinatorial optimization problem given by a cost function, $C^{\rm classical}(\vec{z})$, where $\vec{z}=(z_1,z_2,\dots,z_n)\in\{-1,1\}^n$ is a vector of decision variables. QAOA encodes the decision variables as computational basis states such that $n$-qubit $\ket{\vec{z}}$, with $z_i=\langle\vec{z}|Z_i\ket{\vec{z}}$, encodes~$\vec{z}$. Here, $Z_i$ is a Pauli operator acting at qubit $i$.

QAOA consists of five algorithmic components:
\begin{itemize}[noitemsep]
    \item An initial state, $\ket{\psi_0}$.
    \item A phase separating Hamiltonian, $H_{\mathcal{C}}$, related to the optimization problem by $C^{\rm classical}(\vec{z}) = \langle \vec{z}| H_{\mathcal{C}}|\vec{z} \rangle$. 
    \item A mixing Hamiltonian, $H_{\mathcal{M}}$.
    \item An integer, $p\geq1$, which is typically called the number of QAOA rounds.
    \item A total of $2p$ real-valued numbers that parameterize the time evolution with $H_{\mathcal{C}}$ and $H_{\mathcal{M}}$  for each round. These parameters are referred to as \emph{angles}. For a round number, $i$, we designate the angles as $\gamma_i$ and $\beta_i$, respectively. We introduce a notation $\vec{\gamma} = (\gamma_1,\gamma_2, \dots,\gamma_p)$, and $\vec{\beta}=(\beta_1,\beta_2,\dots,\beta_p)$.  
\end{itemize}

The QAOA protocol consists of alternating simulation of the two non-commuting Hamiltonians, $H_{\mathcal{C}}$ and $H_{\mathcal{M}}$, acting on the initial state, $\ket{\psi_0}$, resulting in the evolved state
\begin{align}
\label{eq:QAOA}
\ket{\vec{\gamma},\vec{\beta}} = & \ e^{- \mi \beta_p H_{\mathcal{M}}}   e^{- \mi \gamma_p H_{\mathcal{C}}} \dots \\
& \dots e^{- \mi \beta_2 H_{\mathcal{M}}}   e^{- \mi \gamma_2 H_{\mathcal{C}}} e^{- \mi \beta_1 H_{\mathcal{M}}}   e^{- \mi \gamma_1 H_{\mathcal{C}}} \ket{\psi_0}. \nonumber
\end{align}
Applications of $H_{\mathcal{C}}$ separate computational  basis states by phases $e^{- \mi \gamma_i C^{\rm classical}(\vec{z})}$,
while $H_{\mathcal{M}}$ produces parameterized interference between states. To generate the classical  optimization problem approximate solutions, that is $\ket{\vec{z}}$ with close to optimal $C^{\rm classical}(\vec{z})$, $\ket{\vec{\gamma},\vec{\beta}}$ is measured in the computational basis. 

If $p$ is sufficiently large and the parameters, $\vec{\beta}$ and $\vec{\gamma}$, are sufficiently well-tuned, this algorithm will sample low energy eigenstates of the underlying classical Hamiltonian $H_{\mathcal{C}}$, which are the computational basis states $\ket{\vec{z}}$. The goal is to sample high-quality approximate solutions of the classical optimization problem, ideally finding a globally optimal solution or solutions. Learning optimal $\vec{\beta}$ and $\vec{\gamma}$ values is computationally challenging in general, but optimal angles are not necessary for QAOA to perform well as the sampling protocol. Nonetheless, learning good parameters is important to get the best results possible, especially on noisy quantum computer hardware. 

QAOA can be treated as an example of a variational quantum algorithm. In this approach, the time-evolved state in Eq.~\eqref{eq:QAOA} is compiled to a parameterized quantum circuit with parameters 
\begin{equation}
\vec{\theta} = (\gamma_1, \gamma_2, \dots,\gamma_p, \beta_1, \beta_2, \dots, \beta_p),
\label{eq:ang}
\end{equation}
and an optimized cost function
\begin{equation}
C(\vec{\theta}) \equiv C(\vec{\gamma}, \vec{\beta}) = \langle \vec{\gamma}, \vec{\beta} | H_{\mathcal{C}}| \vec{\gamma}, \vec{\beta}\rangle.
\label{eq:cost_MaxCut}
\end{equation}
For large enough $p$, the angles resulting in optimal $C(\vec{\gamma,\beta})$ correspond to a high probability of sampling from $\ket{\vec{\gamma},\vec{\beta}}$ states $\ket{\vec{z}}$ with close to optimal $C^{\rm classical}$ , and therefore result in high-quality approximate solutions of the classical optimization problem.

\section{Numerical Results}
\label{sec:numerical_results}

In this section, we present the results of a numerical investigation into the effectiveness of the proposed approach. 
We simulated finite-shot QAOA optimization for Max-Cut problems on random graphs and random Ising models on a heavy-hex lattice. 
We compared our Max-Cut results to the state-of-the-art QAOA optimization method, DARBO. 
In Sec.~\ref{sec:DARBO}, we briefly describe DARBO.
In Sec.~\ref{sec:MaxCut} we present the Max-Cut results and in Sec.~\ref{sec:QAOA_MPS} the random Ising model results.  

\subsection{Reference Method: DARBO}
\label{sec:DARBO}
DARBO~\cite{cheng2024quantum} is a Bayesian optimization algorithm
designed for QAOA.
It uses a Gaussian process with a parametrized Mat\'ern 5/2 kernel as a surrogate function to approximate the cost function by fitting previously evaluated points. 
The kernel parameters are determined by pre-training. 
Rather than fitting all previous points,
it fits only those points inside an intersection of two regions in the parameter space.
The first, called the adaptive trust region, is a hypercube centered on the current best set of parameters.
When the cost function is evaluated for new parameters, this hypercube grows if the value of the cost function consistently improves
(consecutive successes)
and shrinks when it consistently does not
(consecutive failures).
The second, called the adaptive search region,
switches between allowing parameter selection across the whole domain
or a subset of it whenever consecutive failures are detected.
DARBO samples at the maxima of a parametrized acquisition function within the intersection of the trust and search regions.
The acquisition function is chosen heuristically 
to balance sampling in the vicinity of the 
surrogate minima and in high-uncertainty 
parameter space regions.

DARBO has been benchmarked against standard VQA optimization techniques for a Max-Cut problem defined on weighted 3-regular graphs with $n=10-18$ vertices~\cite{cheng2024quantum}. 
In this benchmark, the method was shown to outperform other black-box optimizers including Adam, L-BFGS-B, Nelder-Mead, COBYLA, SPSA, global Bayesian optimization and TURBO. 
DARBO's advantage was especially pronounced when smaller shot numbers were used to evaluate $C(\vec{\theta})$, and has been observed to increase with growing $p$ and $n$. 
Therefore, we treat it as a state-of-the-art reference method.  

\subsection{QAOA for Max-Cut on 3-regular Weighted Graphs}
\label{sec:MaxCut}

We begin with the results for finite-shot QAOA applied to the Max-Cut problem. 
We considered Max-Cut defined for weighted graphs $G=(V,E)$ with a classical Hamiltonian
\begin{equation}
H_{\mathcal{C}} = \sum_{ij} w_{ij} Z_i Z_j. 
\label{eq:H_MaxCut}
\end{equation}
Here $i,j$ index graph vertices ($i, j \in V$) which are connected by edges $\{i,j\} \in E$,   $w_{ij}\in [0,1]$ is the weight associated with an edge and $Z_i$ is a Pauli operator acting at vertex $i$. 
We constructed graph instances by taking $G$ to be 3-regular, with each vertex having three edges and choosing a set of random edge weights. 

More precisely, we used 5 graph instances with $n=|V|=16$ vertices that were used to test DARBO in Ref.~\cite{cheng2024quantum} (These instances can be found in Supplemental Information of Ref.~\cite{cheng2024quantum}).

We constructed the QAOA ansatz by taking $H_{\mathcal{M}} = \sum_i X_i$, for Pauli operators $X_i$, and using an equal weight superposition of computational basis states as an initial state, $\ket{\psi_0}$, defined by $\forall_{i} X_i\ket{\psi_0}=\ket{\psi_0}$. 
Furthermore, we chose $p\in\{2,4,6,8,10\}$. 
We numerically simulated the optimization of Eq.~\eqref{eq:cost_MaxCut} in the absence of hardware noise. 
This matches the reference simulation setup~\cite{cheng2024quantum}.   

To quantify optimization quality we used an approximation ratio defined for a cost function value, $C(\vec{\theta})$, as 
\begin{equation}
r = \frac{C_{\rm max}-C(\vec{\theta})}{C_{\rm max}-C_{\rm min}}, 
\label{eq:app_rat}
\end{equation}
with $C_{\rm max}$ and $C_{\rm min}$ being the maximal and the minimal cost function values, respectively. 
Then, for $C=C_{\rm min}$, we have $r=1$, and, for $C=C_{\rm max}$, we have $r=0$. 
In our case, $C=C_{\rm min}$ ($C=C_{\rm max}$) for the angles that prepare the ground state of $H_{\mathcal{C}}$ ($-H_{\mathcal{C}}$), respectively. 

We compared the approximation ratio of an average optimization run with DARBO. 
The DARBO results were obtained from a repository~\cite{emqaoa}, which contains the data published in Ref.~\cite{cheng2024quantum}. 
We evaluated $C(\vec{\theta})$ using $N_s\in\{200,5000\}$ shots. 
Furthermore, we set the total number of evaluations per run, $N_{\rm init}+N_{\rm it}=1000$, with $N_{\rm init}=50$. 
Consequently, the total number of shots per run was the same for our method and DARBO. 
We chose the $\beta_i$ range to be the full parameter range after taking into account symmetries of the problem~\cite{zhou2020quantum}, i.e. $\forall_i \ \beta_i \in [-\pi/4, \pi/4]$. 
In general, $\gamma_i$ cannot be restricted to a finite range.  
Nevertheless, it has been observed that, in practice, $\gamma_i$ can be bounded while retaining high solution quality~\cite{zhou2020quantum}. 
Here, we have $\forall_i \  \gamma_i\in [-\pi/2, \pi/2]$, the same as for the restricted search space in DARBO. 
The initial points were obtained by random uniform sampling of the parameter ranges. 
For each graph and value of $p$ we conducted $20$ runs, starting each with a different random initialization. 
The same number of randomly initialized runs was used to generate the reference results.

\begin{figure}[t]
        \includegraphics[width=\columnwidth]{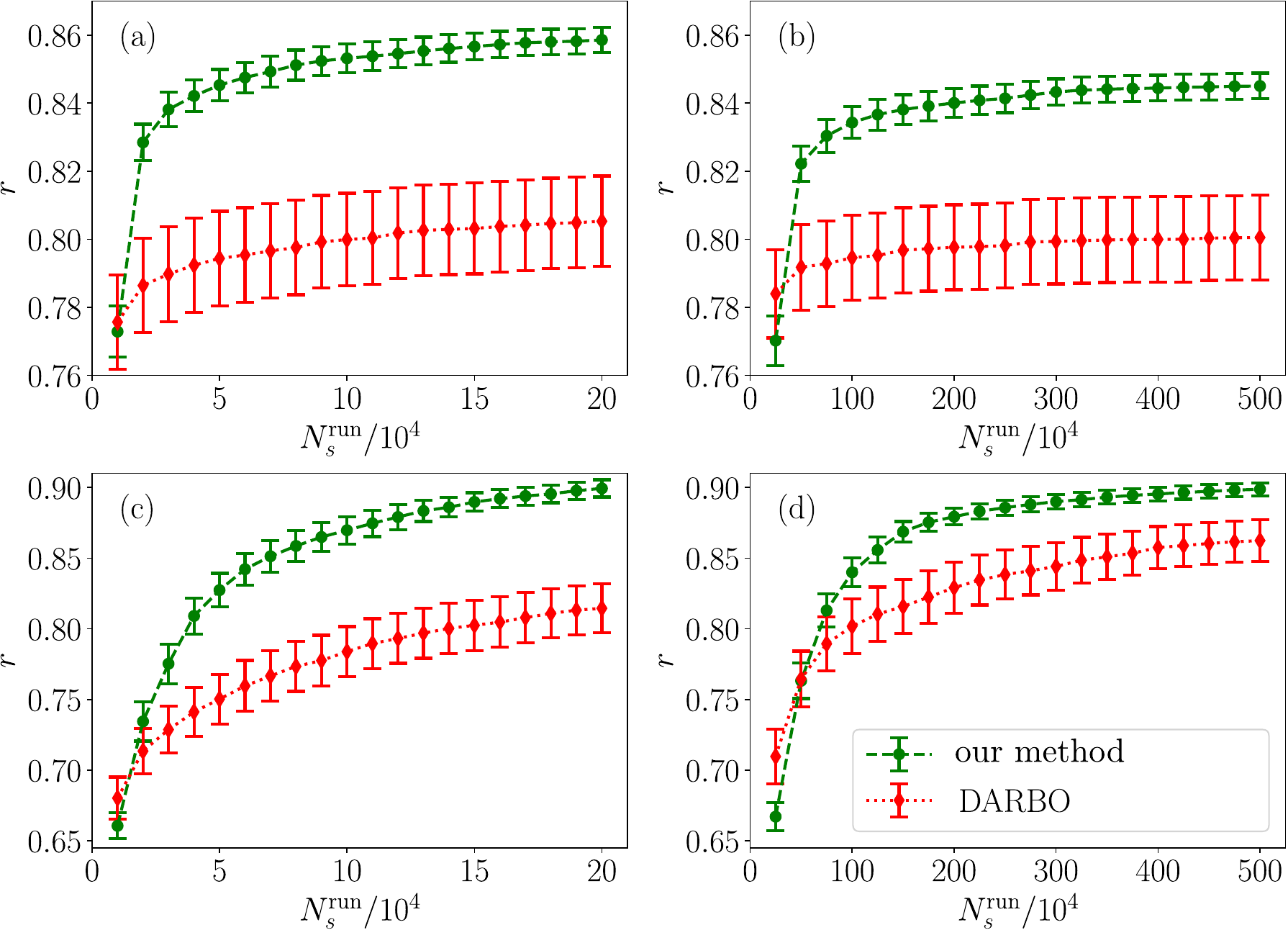}
    \caption{{\bf A comparison of the surrogate-based optimization with DARBO for 3-regular 16-qubit Max-Cut problems.} Approximation ratios, Eq.~\eqref{eq:app_rat}, of the best cost function values found by our method (green) and DARBO (red) runs within the first $N_s^{\rm run}$ shots of the run. In the upper (lower) row we show results for $p=2$ ($p=10$), respectively. The optimization was performed in the presence of numerically simulated shot noise with $N_s=200$ shots per cost function evaluation in the left column and $N_s=5000$ in the right one. The markers represent $r$ averaged over 5 random graphs and 20 randomly initialized runs per graph. The error bars are the mean standard deviations multiplied by a factor of 2, and correspond to a $95\%$ confidence interval. 
    }
    \label{fig:MaxCut}
\end{figure}

In Fig.~\ref{fig:MaxCut}, we compare results for $p=2$ and $p=10$.  
These are the smallest and largest $p$ considered in Ref.~\cite{cheng2024quantum}. 
We present the approximation ratio of the best cost function evaluated so far, $ {\rm min}\, \{C(\vec{\theta}) \,|\, \vec{\theta} \in  \Theta \}$, versus a run shot cost 
\begin{equation}
N_s^{\rm run} = N_s(N_{\rm init} + i).
\label{eq:cost_tot}
\end{equation}
Here, $i$ indexes optimization iteration number, as described in detail in Sec.~\ref{sec:alg}. 
In Fig.~\ref{fig:MaxCut}, we plot the approximation ratio, $r$ averaged over all runs per graph and graph instances.
The error bars are the mean standard deviation multiplied by a factor of~$2$.  
We find that, apart from the first iterations, our approach outperforms DARBO, resulting in higher $r$ for $N_s^{\rm run}>10^4$ ($N_s^{\rm run}>5\cdot 10^5$), when $N_s=200$ ($N_s=5000$), respectively. 
The performance gap is especially pronounced for $N_s=200$, where for $N_s^{\rm run}=10^5$ and $p=2$ our approach obtained $r=0.859(2)$ versus the $r=0.784(7)$ achieved by DARBO. 
Similarly, for $p=10$ and the same $N_s^{\rm run}$, our method and DARBO achieved $r=0.899(3)$ and $r=0.795(9)$ respectively. 
We note that the shot budgets considered here ($N_s^{\rm run}<5\cdot10^6$) match what is currently available through cloud access to quantum computers. 

\begin{figure}[t!]
    \includegraphics[width=0.99\columnwidth]{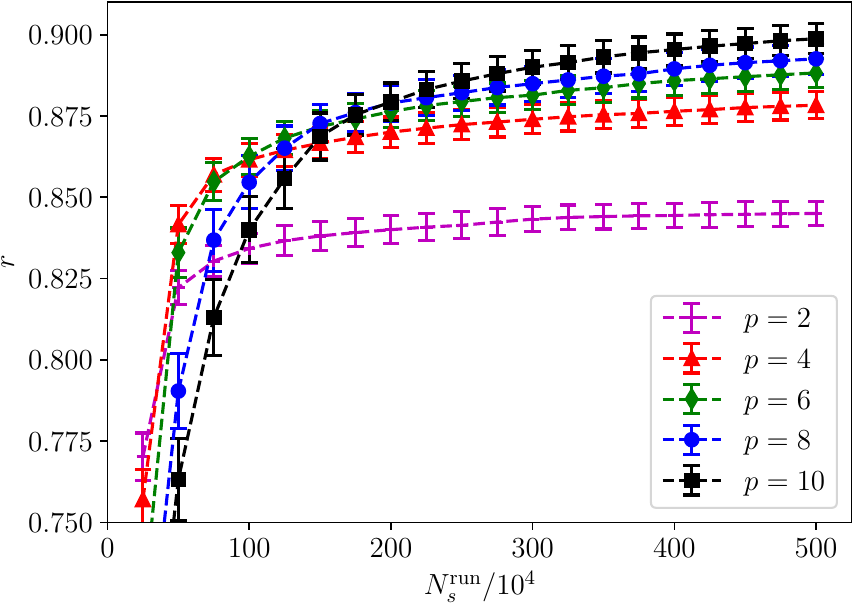}
    \caption{ {\bf Surrogate-based optimization performance with increasing number of  QAOA rounds for 3-regular 16-qubit  Max-Cut problems.} The approximation ratio of the best cost, $C(\vec{\theta})$, found during the first  $N_s^{\rm run}$ shots of an optimization run is plotted for multiple QAOA round numbers $p=2$ (magenta), $p=4$ (red), $p=6$ (green), $p=8$ (blue), $p=10$ (black). Here the number of shots per $C(\vec{\theta})$ evaluation was $N_s=5000$.  The choice of graphs, optimization initialization and hyperparameters were the same as in  Fig.~\ref{fig:MaxCut}. The markers denote the mean values obtained by averaging over 5 graphs and 20 runs per graph, and the error bars were computed in the same way as in Fig.~\ref{fig:MaxCut}. }
    \label{fig:MaxCut_pconv}
\end{figure}

Next, we analyze the quality of our surrogate-based optimization method as a function of the number of QAOA rounds, comparing $p\in\{2,4,6,8,10\}$ runs. 
We present these results in Fig.~\ref{fig:MaxCut_pconv}, obtained using the same 5~graphs, initialization method, optimization hyperparameters, and number of runs per graph as for the DARBO comparison. 
We find with $N_s=5000$ that, for large enough $N_s^{\rm tot} \ge 2\cdot10^6$, the results systematically improve with increasing $p$.  
Therefore, despite using the limited shot budget available with current quantum devices, the method is capable of harnessing the increase in QAOA ansatz expressibility with increasing $p$. 

Additionally, in Appendix~\ref{app:MaxCut_add_res} we present the values of $C(\vec{\theta})$ computed with infinite shots and the best angles found after the finite-shot optimization runs.
This is to both understand the effect of finite shot noise on the estimates of $C(\vec{\theta})$ used in Figs.~\ref{fig:MaxCut},~\ref{fig:MaxCut_pconv}, and to more robustly quantify the quality of the angles found using finite-shot optimization.
In this case, we find that our approach's advantage over DARBO, and the convergence with increasing $p$, remain very similar to results shown in Figs.~\ref{fig:MaxCut},~\ref{fig:MaxCut_pconv}. 

\subsection{QAOA for Random Ising Models on the 127-qubit Heavy-Hex Hardware Graph}
\label{sec:QAOA_MPS}

Next, we considered optimization of the QAOA ansatz defined in Eq.~(\ref{eq:QAOA}) with $p=3$ for a 127-qubit graph compatible with 133-qubit \texttt{ibm\_torino} quantum processor connectivity. 
\texttt{ibm\_torino} is a superconducting qubit quantum computer that has a sparse 2-qubit gate interaction graph, generally referred to as heavy-hex connectivity~\cite{chamberland2020topological}. 
We focused on 5 instances of a random Ising model with a problem Hamiltonian
\begin{equation}
H_{\mathcal{C}} = \sum_i d_i Z_i + \sum_{\langle i,j \rangle} d^{(2)}_{i,j} Z_i Z_j + \sum_i d^{(3)}_{i,j,k}  Z_i Z_{j} Z_{k}. 
\label{eq:Hdev}
\end{equation}
Here, in the second term, we sum over edges of the graph, and the three-body terms occur for a subset of graph nodes indexed by $i$ that have exactly two nearest-neighbors $j,k$, see  Refs.~\cite{pelofske2024short, pelofske2023quantum, pelofske2023scaling} for details. The coefficients $d_i, d^{(2)}_{i,j}, d^{(3)}_{i,j,k
} \in \{-1,1\}$ are chosen uniformly at random with probability $0.5$.

This class of optimization problem has been designed to be highly hardware-compatible with the IBM processors that share this heavy-hex graph structure. 
In particular, despite including the three-body terms in $H_{\mathcal{C}}$, a single QAOA round can be implemented using a circuit with CNOT gate depth of $6$. 
The graph sparsity also means that these optimization problems are relatively easy to sample using heuristic methods such as simulated annealing. 
Moreover, they can be solved exactly using optimization software such as CPLEX~\cite{pelofske2024short, pelofske2023quantum, pelofske2023scaling}. 
QAOA performance for these problems has been investigated using 127-qubit IBM quantum processors~\cite{pelofske2023scaling} using angles learned with a technique known as parameter transfer. 
In a variety of combinatorial optimization problem instances, it has been observed that QAOA angles seem to perform similarly for many different problem instances that are all from a similar class of problem type (i.e. similar graph structure)~\cite{galda2021transferability, akshay2021parameter, brandao2018fixed, farhi2022quantum, galda2023similaritybased, shaydulin2023parameter}. 
This property is typically called \emph{parameter concentration} or \emph{parameter transfer}. Ref.~\cite{pelofske2023scaling} used this strategy to train QAOA on a $16$-qubit heavy-hex Ising model instance with Hamiltonian in Eq.~\eqref{eq:Hdev}, and to apply the learned angles to larger problem instances with the same Hamiltonian. 
In this section, we investigate the capability of our surrogate-based optimization approach to improve on the parameter transfer angles found in Ref.~\cite{pelofske2023scaling}. 
We will refer to these parameter transfer angles as heuristic angles.

We numerically simulated finite-shot optimization for 5 random Hamiltonian instances, $p=3$ and with $N_s=1000$ shots per $C(\vec{\theta})$ evaluation using a Matrix Product State (MPS) simulator~\cite{fannes1992finitely,pelofske2023scaling}. 
In our simulations we assumed no hardware noise. 
The set of initial parameters $\Theta_{\rm init}$ is built of the parameter transfer angles
$\vec{\theta}_{\rm heur}^{p=3}$, and $19$ random vectors of parameter $\vec{\theta}$ (\ref{eq:ang}) choices. Each random choice is obtained by uniform random sampling of a full parameter range for each of $2p$ $\vec{\theta}$ elements. As the cost function in Eq.~\eqref{eq:cost_MaxCut} is invariant under shifts of $\gamma_i$ and $\beta_i$ by $\pi$, we choose the full parameter ranges to be $\gamma_i \in [-\pi/2,\pi/2),\, \beta_i \in [-\pi/2,\pi/2)$.   
The parameter transfer angles were found in Ref.~\cite{pelofske2023scaling} by numerically optimizing small $16$ qubit instances of the random Ising model defined in Eq.~\eqref{eq:Hdev} on a heavy-hex graph, and are listed in Table ~\ref{tab:heuristic_angles}. 
Furthermore, we used $N_{\rm it}=80$ optimization iterations during training. Our hyperparameter choices resulted in a total run shot cost of $10^5$, which is accessible with cloud-based access to current IBM quantum computers.  

\begin{figure}[t]
    \includegraphics[width=0.9\columnwidth]{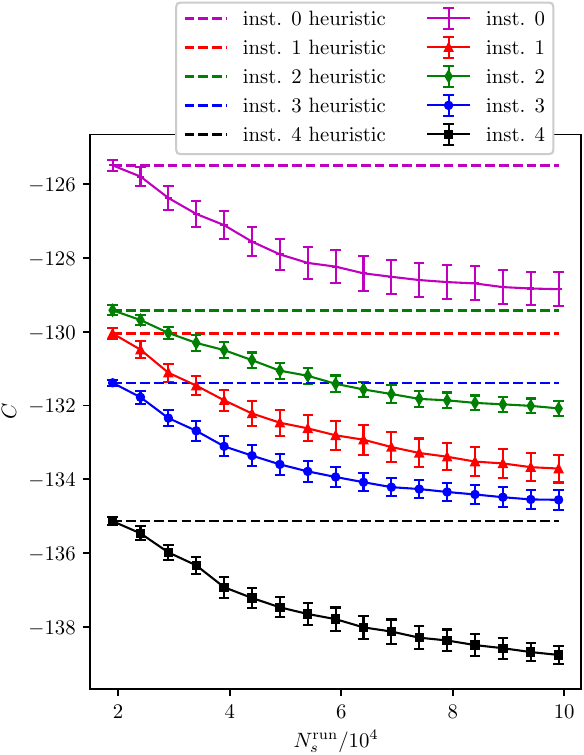}
    \caption{ {\bf Numerical Matrix Product State optimization for $p=3$ QAOA and 127-qubit heavy-hex Ising models.} Here we plot the best cost $C(\vec{\theta})$ (energy) found during the first $N_s^{\rm run}$ run shots (solid lines) for 5 instances of the model coefficients (\ref{eq:Hdev}) and $N_s=1000$ shots per $C(\vec{\theta})$ evaluation. The dashed vertical lines indicate the  cost function  for  the parameter transfer heuristic angles $\vec{\theta}_{\rm heur}^{p=3}$, which are included as one of  $N_{\rm init}=20$ initial angles. The remaining initial parameters are chosen randomly. The markers show the mean values obtained by averaging over $32$ runs started with different choices of the random initial angles.  
    The error bars are the mean standard deviations multiplied by a factor of $2$, which correspond to $95\%$ confidence intervals.  
    }
    \label{fig:hhex_MPS}
\end{figure}

\begin{table}[ht]
    \centering
    \begin{tabular}{|c|c|}
        \hline
         $\vec{\beta}_{\rm heur}^{p=3}$ & $(0.50502,  0.35713,  0.19264)$ \\ \hline
         $\vec{\gamma}_{\rm heur}^{p=3}$ & $(-0.14264, -0.26589, -0.34195)$ \\ \hline
        $\vec{\beta}_{\rm heur}^{p=4}$ & $(0.54321,0.41806,0.28615,0.16041)$ \\ \hline 
        $\vec{\gamma}_{\rm heur}^{p=4}$ &  $(-0.12077,-0.22360,-0.29902,-0.35329)$ \\ \hline
        $\vec{\beta}_{\rm heur}^{p=5}$  & $(0.53822, 0.44776, 0.32923, 0.23056, 0.12587)$ \\ \hline
        $\vec{\gamma}_{\rm heur}^{p=5}$  & $(-0.11764, -0.19946, -0.268736, -0.321586, -0.34583)$ \\ \hline
    \end{tabular}
    \caption{ { \bf Heuristic (parameter transfer) QAOA angles for $p\in\{3,4,5\}$ rounds on random-coefficient Ising Hamiltonians with heavy-hex connectivity structure, $\vec{\theta}_{\rm heur}=(\vec{\gamma}_{\rm heur},\vec{\beta}_{\rm heur})$.} The angles are obtained by a numerical optimization on a 16-qubit graph in Ref.~\cite{pelofske2023scaling}. The heuristic angles are included in sets of initial angles $\Theta_{\rm init}$ for the surrogate-based optimization of the 127-qubit instances. The angles for $p=4$ and $p=5$ are used in Sec.~\ref{sec:dev_opt}.}
    \label{tab:heuristic_angles}
\end{table}

For multiple types of classical problem Hamiltonian, it has been shown that the parameter transfer angles found by optimization of similar problems work very well for most other problem instances in that class \cite{galda2021transferability, akshay2021parameter, brandao2018fixed, farhi2022quantum, galda2023similaritybased, shaydulin2023parameter}. 
In particular, Ref.~\cite{pelofske2023scaling} has shown this for the heuristic angles and the heavy-hex Ising model problem instances considered here. 
Moreover, it has been shown that, for low-enough $p$, QAOA cost function landscapes frequently have multiple suboptimal local minima which limits ansatz trainability~\cite{rajakumar2024trainability}. 
Therefore, we expect that the inclusion of $\vec{\theta}_{\rm heur}^{p=3}$ among the initial angles may speed up the convergence of our approach towards angles with low-cost function values. 
Furthermore, we expect that for this case a fully random initialization might reduce optimization shot-efficiency, since many calls to the true cost function may be required to learn a complex cost landscape from scratch. 

\begin{figure*}[t]
    \includegraphics[width=0.99\textwidth]{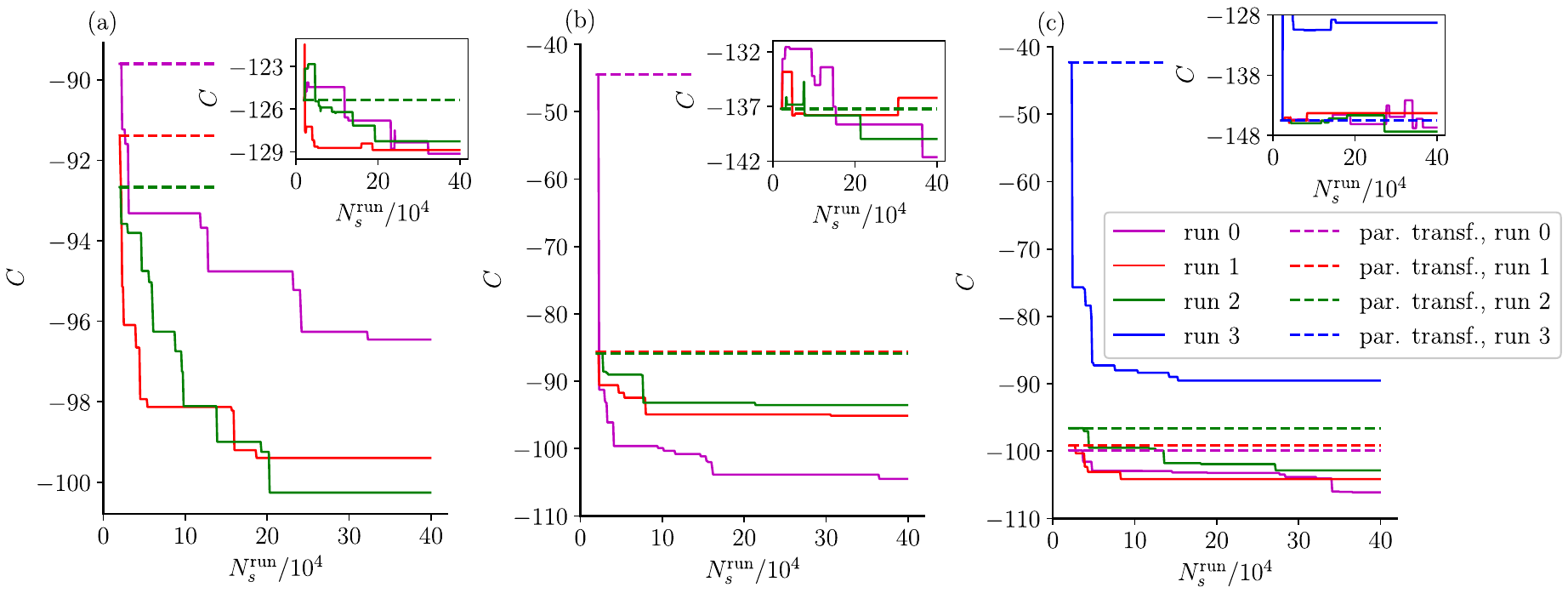}
    \caption{ { \bf Surrogate-based optimization on the \texttt{ibm\_torino} quantum computer for $p\in\{3, 4, 5\}$ QAOA applied to an instance of a 127-qubit heavy-hex random Ising model.} Here, (a) depicts $p=3$, (b) $p=4$, and  (c) $p=5$ optimization results. In the main plots, the solid lines are the best device cost function estimates $C(\vec{\theta})$, for $N_s=1000$, found within the first  $N_s^{\rm run}$ shots for several runs. The solid lines in the insets show the infinite-shot MPS cost function, which is not affected by hardware noise, for the same angles as in the main plots. The dashed horizontal lines shown in both cases are the cost function estimates for the parameter transfer angles from Ref.~\cite{pelofske2023scaling}. The parameter transfer angles are included in the runs' initial angles $\Theta_{\rm init}$.
    }
    \label{fig:torino}
\end{figure*}

In Fig.~\ref{fig:hhex_MPS}, we show the best cost function found, ${\rm min}\, \{C(\vec{\theta})\, | \, \vec{\theta} \in \Theta \}$, versus a run shot cost $N_s^{\rm run}$ of Eq.~(\ref{eq:cost_tot}), for each of the $5$ Hamiltonian instances. 
The cost was averaged over $32$ runs, each started with different choices of random angles in $\Theta_{\rm init}$. 
Furthermore, we compare the optimized cost function to its value at the parameter transfer angles. 
As expected, for all Hamiltonian instances, we find that the randomly sampled initial points do not improve upon $C(\vec{\theta}_{\rm heur}^{p=3})$, as shown by the first evaluated points on the learning curve having the same energy as the parameter transfer angles for each problem instance. 
In contrast, the surrogate-based optimization systematically improves upon the parameter transfer angles. 
This improvement is clearly visible for all instances, even for $N_s^{\rm run}$ as small as $3\cdot10^4$.

To provide further evidence for this conclusion, we present infinite-shot cost function values in Appendix ~\ref{sec:heavy_hex_add_res}.
This is in order to more robustly quantify the cost improvement for the angles found by our method. 
That is, we determined the best angles, ${\rm argmin}\, \{C(\vec{\theta})\, | \, \vec{\theta} \in \Theta \}$, using $N_s=1000$ cost function estimates from the optimization runs (the same as in the main text). 
Subsequently, we computed infinite-shot $C(\vec{\theta})$ for these angles.
The observed behavior of $C(\vec{\theta})$ is very similar to the finite shot case, as shown in Fig.~\ref{fig:hhex_MPS}.  
Additionally, in Appendix~\ref{sec:heavy_hex_add_res}, we show that these cost function estimates appear to be converged in the MPS bond dimension, $D$, the refinement parameter of the numerical simulations.  
 
\section{Hardware Implementation} 
\label{sec:dev_opt}

\begin{figure}[t]
    \includegraphics[trim=0 4cm 0 6cm,width=0.98\columnwidth]{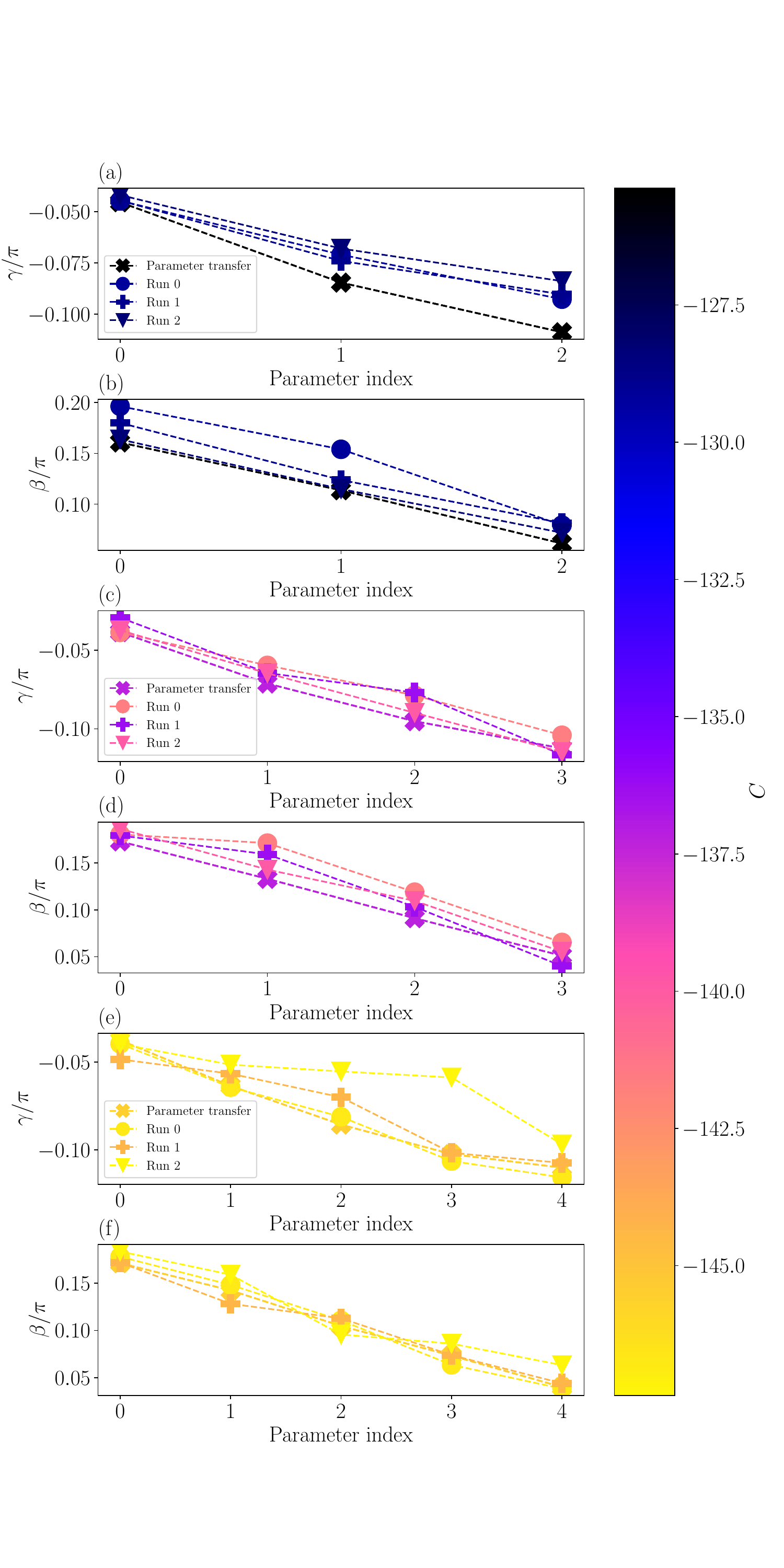}
    \caption{{\bf Comparison of the initial parameter transfer angles  $\vec{\theta}_{\rm heur}$ with the angles learned from noisy device optimization, $\vec{\theta}_{\rm opt}$, having been initialized with the parameter transfer angles.}  The plots show the final angles from runs 0, 1, 2 in Fig.~\ref{fig:torino}, and the parameter transfer runs from Table~\ref{tab:heuristic_angles}. 
   Panels (a) and (b) show results for $p=3$ QAOA rounds, (c) and (d) plot $p=4$ results, and (e, f) contain $p=5$ angles.
    Parameter index $i$, see (\ref{eq:QAOA}), specifies to which QAOA layer the angles belong to. 
    The color bar is the QAOA cost function evaluated at the angles shown using an MPS simulator in the infinite-shot limit.}
    \label{fig:ang_evol}
\end{figure}

We performed surrogate-based optimization for an instance of a 127-qubit heavy-hex Ising model (instance $0$ from Fig.~\ref{fig:hhex_MPS}) with a $p\in\{3,4,5\}$ QAOA ansatz on the \texttt{ibm\_torino} quantum computer, using $N_s=1000$ shots for each cost function evaluation. 
We chose $\Theta_{\rm init}$, and the angle bounds in the same way as for the MPS optimization in Sec.~\ref{sec:QAOA_MPS}. 
More precisely, $\Theta_{\rm init}$ was the union of 19 random angles and $\vec{\theta}_{\rm heur}$ from Table~\ref{tab:heuristic_angles}.
During optimization the QAOA angles were bound to lie within $[-\pi/2,\pi/2)$.
We set the iteration number $N_{\rm it} = 380$.
The optimization hyperparameters are presented in more detail in Appendix~\ref{app:surr_config}.
For each value of $p$, we conducted $3$ or $4$ optimization runs, and plot the best device cost function estimates versus shot number (\ref{eq:cost_tot}) in Fig.~\ref{fig:torino}. 
Additionally, to benchmark angle improvement, we evaluated MPS infinite-shot cost values for the angles with the best device $C(\vec{\theta})$, and show the results in the insets of Fig.~\ref{fig:torino}. 
The MPS cost values were not affected by the hardware noise, and were converged in the MPS bond dimension $D$, as discussed in Appendix~\ref{app:MPS_hard}. Therefore, the MPS cost values reliably quantify optimized angle quality. 

We observe that our surrogate-based approach systematically improves on the initial cost function values.
For all runs, a significant part of the total improvement occurred for $N_s^{\rm tot} < 5 \cdot 10^{4}$. 
Furthermore, for $p\in \{3,4\}$ the vast majority of the gain happened for $N_s^{\rm tot} < 2 \cdot 10^{5}$. 
This demonstrates that the shot-efficiency of the surrogate-based approach enables on-device training with limited, cloud-based access to the quantum hardware. 

The impact of the hardware noise is clearly demonstrated by values of $C(\vec{\theta}_{\rm heur}^{p=3})$ obtained from the device, $\{-92.7,-91.4,-89.6\}$, that are larger than the MPS value of $-125.4$. 
Furthermore, for $p=4$, we see a large discrepancy of the device cost values $C(\vec{\theta}_{\rm heur}^{p=4}) \in \{-102,-101,-44.5\}$, which is unlikely to be caused by finite $N_s$ effects. 
A more likely explanation is the time variability of hardware noise. We see a similar effect for the $p=5$ runs.
In our MPS benchmarks, for $p=3$ we observe that all 3 runs converged to the cost function values typical for MPS optimization, see Fig.~\ref{fig:hhex_MPS} for comparison, improving upon the heuristic MPS cost function, and showcasing the noise-resilience of the QAOA optimization for shallow-enough circuits. 
In contrast, for the deeper $p=4$ and $p=5$ circuits, the final MPS cost $C_{\rm opt}$ only improved on the parameter transfer angle cost for 2 out of 3 and 2 out of 4 runs, respectively. 
Therefore, for $p=4$ and $p=5$, the noise resilience of the on-device QAOA optimization is significantly reduced. 
Overall, the results demonstrate that despite significant hardware noise and device instability, our approach is able to find good QAOA angles and improve on already known QAOA angles. 

An overview of the optimized QAOA parameters $\vec{\theta}_{opt}$, which are obtained at the end of each run, is shown in Fig.~\ref{fig:ang_evol}. 
This plot compares $\vec{\theta}_{\rm opt}$ and
$\vec{\theta}_{\rm heur}$, correlating the parameter values with their corresponding infinite-shot MPS cost values.
We observe that for all $p$, $|\gamma_i|$ increases with increasing parameter index $i$, while $|\beta_i|$ decreases.
At $p=3$, the newly learned values of $\gamma_i$ and $\beta_i$ are typically larger than for the parameter transfer angles. 
For noise-resilience-breaking $p\in\{4,5\}$ it is less clear whether there are patterns distinguishing $\vec{\theta}_{\rm opt}$ and
$\vec{\theta}_{\rm heur}$. 

Finally, we investigated whether the improved QAOA angles learned by our surrogate-based method could be successfully transferred to larger problem instances.
This was done for $p=3$, by comparing cost values for the QAOA parameters learned after 127-qubit device optimization $C(\vec{\theta}_{\rm opt})$ with $C(\vec{\theta}_{\rm heur}^{p=3})$, for 133 and 156-qubit random Ising model instances defined using Eq.~(\ref{eq:Hdev}).
The cost function for both instances was evaluated using 8192 shots on the 156-qubit \texttt{ibm\_marrakesh} processor and in the infinite-shot limit using an MPS simulator. 
The cost values are presented in Table~\ref{tab:param_transfer}.
We see that the angles learned using a noisy quantum device provide a consistent improvement over the previously known angles for these new, larger problem instances. These results further indicate that our method is capable of improving estimates of good QAOA angles. 

\begin{table}[t]
\centering
\begin{tabular}{|c|c|c|c|c|c|c|}
\hline
           $n$ & \multicolumn{3}{|c|}{133} & \multicolumn{3}{c|}{156} \\ \hline
           Eval. nb. & 0 & 1 & 2 & 0 & 1 & 2 \\ \hline
Device $C(\vec{\theta}_{\rm heur}^{p=3})$  & -94.8       & -95.2      & -94.6     & -115       & -110       & -109    \\ \hline MPS $C(\vec{\theta}_{\rm heur}^{p=3})$ & -139      & --       & --     &  -163      &  --     &  --    \\ \hline
Device $C(\vec{\theta}_{\rm opt})$     & -97.7      & -97.4      & -97.3       &  -121       & -116      & -119    \\ \hline
MPS $C(\vec{\theta}_{\rm opt})$      & -143      & -144      & -143        &  -173      & -171 & -169      \\ \hline
\end{tabular}
\caption{ {\bf Parameter transfer of the \texttt{ibm\_torino} quantum computer $n=127, \,p=3$ optimization runs' final angles $\vec{\theta}_{\rm opt}$ to larger (unseen) 133, and 156-qubit random heavy-hex Ising model (eq. \ref{eq:Hdev}) instances.} 
Here, the QAOA cost $C(\vec{\theta})$ was evaluated 
on the \texttt{ibm\_marrakesh} quantum computer with $N_s=8192$ shots. Furthermore, it was also evaluated with $D=2048$ Matrix Product States in the infinite shot limit. We compare the cost obtained for the optimized angles $C(\vec{\theta}_{\rm opt})$ with the cost for the heuristic $\vec{\theta}_{\rm heur}^{p=3}$ angles from Table~\ref{tab:heuristic_angles}.   
In the case of $C(\vec{\theta}_{\rm opt})$, the evaluation number from this table is the same as the optimization run number, used in Figs.~\ref{fig:torino},~\ref{fig:ang_evol}, of a run that produced $\vec{\theta}_{\rm opt}$. In the case of $C(\vec{\theta}_{\rm heur}^{p=3})$ and the device cost estimates, the evaluation number indexes independent evaluations of the cost function for the same angles. As the MPS cost function was evaluated in the infinite-shot limit, and was not subject to hardware noise, it was evaluated once for each graph instance.}
\label{tab:param_transfer}
\end{table}

\section{Discussion}
\label{sec:conclusions}

In this work, we propose a surrogate-based approach that we show is well-suited for on-device optimization of parameterized quantum circuits. 
In our approach, an initial surrogate is built by sampling the true cost function for a set of sparse random parameters.
These random parameters may be supplemented by heuristically chosen parameter values. 
The initial surrogate is then iteratively refined by adaptive sampling at the surrogate optima, with the goal of efficiently improving surrogate accuracy in their vicinity. 

We test the proposed approach by applying it to QAOA. First, we demonstrate that we outperform a state-of-the-art method in QAOA optimization for Max-Cut problems defined on 16-qubit 3-regular random graphs for realistic shot budgets and fully random surrogate initialization. 
We then demonstrate that a noisy quantum processor can be used to systematically improve the QAOA cost compared to the cost for the previously best-known parameter transfer angles for 127-qubit instances of random Ising models on a heavy-hex graph. 
Furthermore, we show that the improved QAOA angles also outperform the initial parameter transfer angles on unseen 133 and 156-qubit problem instances.

These demonstrations are for numerically simulated $p=3$ optimization and on-device $p=3,4,5$ optimization.
Additionally, using a tensor network numerical method, we demonstrate the noise resilience of the $p=3$ hardware optimization and angle improvement on unseen problem instances.

We expect that the shot-efficiency of the proposed approach can be enhanced further. 
In particular, in this work we terminate the adaptive sampling after $N_{\rm it}$ iterations. 
State-of-the-art surrogate methods frequently use composite termination criteria instead, for example taking into account the cost function change in recent iterations. 
Such criteria allow for a faster termination of the optimization if the cost does not change significantly. 
Moreover, imposing physical constraints on the surrogates, like periodicity of the cost as a function of QAOA angles, may further reduce the number of the hardware calls by constructing a surrogate which more closely reflects the true cost.   

Another direction to improve result quality, and ultimately the efficiency of the approach, is sampling strategies which target exploration of sparsely sampled parts of the parameter space. 
Our algorithm attempts to efficiently locate the best local optimum detected out of a set whose presence is inferred from the initial data points. 
However, this approach may not be efficient enough in the presence of many local optima, especially when no heuristic information about their approximate location is available. 
In such circumstances, we expect that strategies involving multiple optimization runs, with new runs starting in the least probed regions of parameter space, may be necessary. 

A related issue emerges when considering parts of the cost landscape with vanishing gradients.
In the presence of barren plateaus, cost function regions with significant variance occupy vanishingly small portions of parameter space~\cite{arrasmith2021equivalence}. 
Nevertheless, heuristic strategies to approximately locate such regions have been proposed, enabling optimization~\cite{puig2024variational}. 
However, even with such heuristic information, the presence of barren plateaus will have a negative impact on the efficiency of sampling strategies which explore the whole parameter space. 
Therefore, to preserve the efficiency of our method, modifications to these strategies to target the non-vanishing gradient regions may be very beneficial.

To correctly identify the barren plateau regions, one needs to handle cost variations caused by finite shot noise. 
In the region of a barren plateau these variations obscure cost changes due to varying parameters. 
Our current approach does not take into account finite shot effects on cost estimates. 
Consequently, directly accounting for shot noise is an interesting topic to explore. 
This development could improve method efficiency even in the absence of barren plateaus. 

While this work concerns continuous parameter optimization for a fixed ansatz, variational approaches with discrete parameters that modify ansatz structure have been proposed~\cite{grimsley2019adaptive}. 
An extension of this work to such cases would be of interest. 
Moreover, as we target QAOA as our proof-of-principle application, a natural continuation of this research is applying the method to other VQAs, for example variational quantum eigensolvers targeting strongly-correlated systems in quantum chemistry~\cite{kandala2017hardware}. 

One potential use of our surrogate-based approach is improving the efficiency of PQC optimization with numerically-expensive and accurate classical approaches like state-of-the-art tensor network~\cite{tindall2023efficient, patra2023efficient}, Pauli propagation~\cite{rudolph2023classical, nemkov2023fourier} or symmetry-exploiting approaches~\cite{anschuetz2022efficient,goh2023lie}. 
Our approach can be applied to minimize the number of cost evaluations by such numerical methods to extend the range of optimization problems to which they are applicable. 
Another future research direction is the usage of such numerical methods to supplement device-derived data by providing accurate results for special classically simulable parameter values, like QAOA angles corresponding to near-Clifford circuits~\cite{beguvsic2023simulating}. 

Finally, a future research direction is combining the surrogate-based approach to on-device learning with error mitigation methods~\cite{cai2022quantum, barron2024provable, sack2024large} and strategies to reduce the effects of hardware-drift. 
A pertinent question is to what extent these methods can improve the noise-resilience of optimization. 
In particular, such improvements may enable successful and reliable on-device QAOA optimization for $p>3$, for which we observe the breakdown of noise resilience. 
Another relevant question is to what extent error mitigation methods are sufficient to reliably judge the noise resilience of optimization~\cite{wang2021can}, as for deep enough ans{\"a}tze, the MPS numerical methods applied in this manuscript are expected to fail. 

\section*{Acknowledgments}
\label{sec:acknowledgments}
We thank Marco Cerezo for helpful conversations. 
This work was supported by the U.S. Department of Energy through the Los Alamos National Laboratory. Los Alamos National Laboratory is operated by Triad National Security, LLC, for the National Nuclear Security Administration of U.S. Department of Energy (Contract No. 89233218CNA000001). The research for this publication was supported by a grant from the Priority Research Area DigiWorld under the Strategic Programme Excellence Initiative at Jagiellonian University. PC acknowledges support  by  the National Science Centre (NCN), Poland under project 2022/47/D/ST2/03393.
This material is based upon work supported by the U.S. Department of Energy, Office of Science, National Quantum Information Science Research Centers, Quantum Science Center (QSC). ATS acknowledges initial support from the LANL ASC Beyond Moore’s Law project and subsequent support from the QSC. MM acknowledges support by the Uncertainty Quantification Foundation under the Statistical Learning program. The Uncertainty Quantification Foundation is a nonprofit dedicated to the advancement of predictive science through research, education, and the development and dissemination of advanced technologies. EP acknowledges support from the NNSA's ASC Beyond Moore's Law Program at LANL. This research used resources provided by the Los Alamos National Laboratory Institutional Computing Program. We acknowledge the use of IBM Quantum services for this work. The views expressed are those of the authors, and do not reflect the official policy or position of IBM or the IBM Quantum team. 

\appendix

\section{Additional Numerical Results}

Here we gather additional numerical data complementing results presented in the main text. 

\subsection{QAOA for 3-regular Max-Cut Weighted Graphs}
\label{app:MaxCut_add_res}

\begin{figure}[htb]
        \includegraphics[width=\columnwidth]{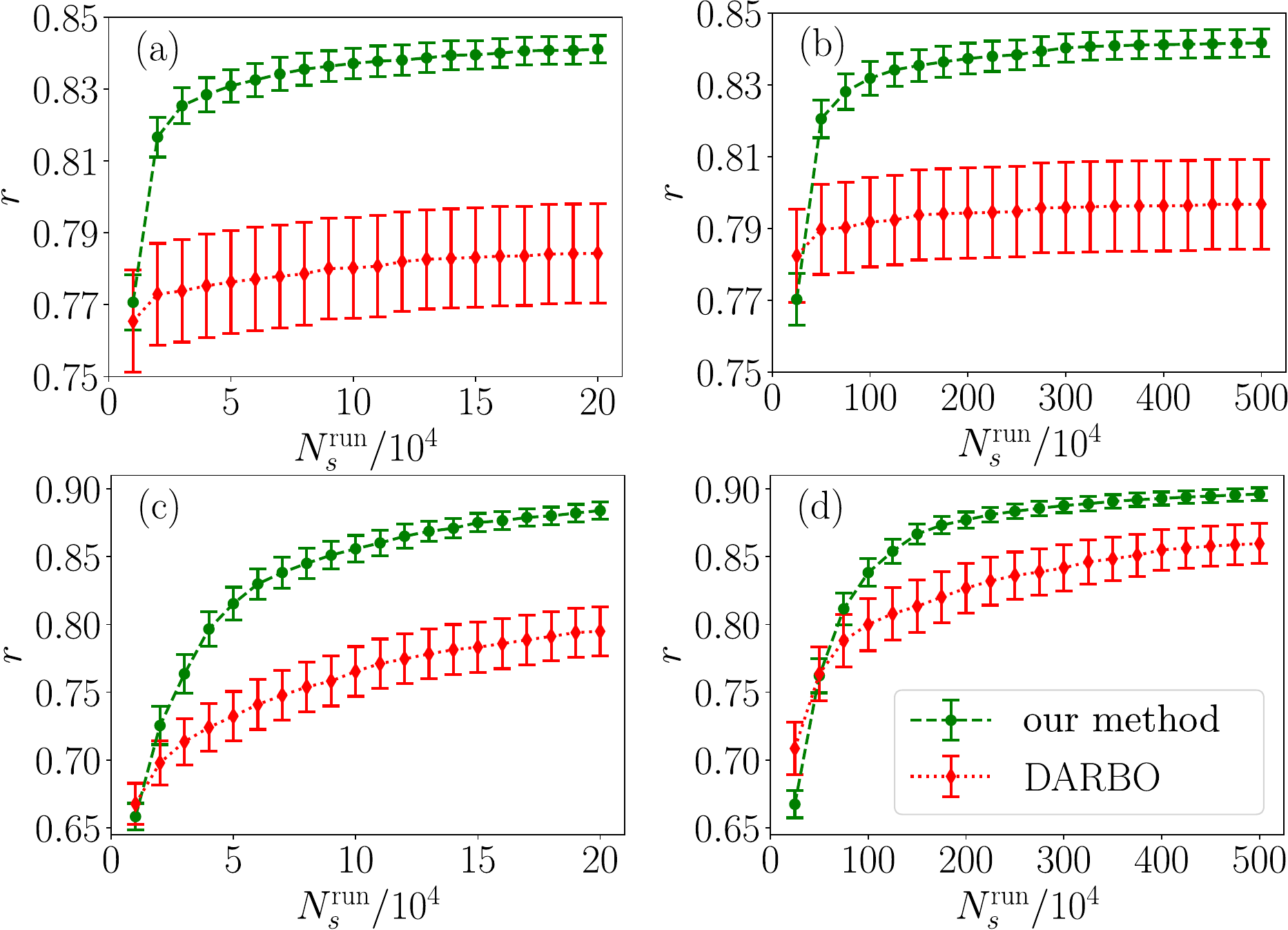}
    \caption{ {\bf A comparison of DARBO and the surrogate-based optimization for the 3-regular 16-qubit Max-Cut problems using the infinite-shot cost function estimates.} We compare the performance of the optimization runs from Fig.~\ref{fig:MaxCut} using approximation ratios (\ref{eq:app_rat}) computed with the infinite-shot cost function values, instead of the finite-shot estimates used in Fig.~\ref{fig:MaxCut}. In both cases, the approximation ratios of  $C(\vec{\theta})$ are computed for the same best angles $\vec{\theta}$, which are found by optimization runs minimizing finite-shot $C(\vec{\theta})$ estimates.   In~(a,b) $p=2$ results, and in~(c,d) $p=10$ results. The $r$ values for $N_s=200$ optimization are gathered in panels~(a,c). We show the results for $N_s=5000$ in plots~(b,d).  The averaging, and the error bar computation are performed the same as in  Fig.~\ref{fig:MaxCut}. }
    \label{fig:MaxCut_shot_inf}
\end{figure}

Here to clarify effects of finite $N_s$ on the QAOA cost function estimates used by the optimization, and to quantify more robustly the quality of the optimized angles, we plot the approximation ratios (\ref{eq:app_rat}) for the best angles found within the first $N_s^{\rm tot}$ shots of an optimization run, ${\rm argmin} \, \{C(\vec{\theta})\, |\,  \vec{\theta} \in \Theta \}$, using infinite-shot cost function $C(\vec{\theta})$ values. That is, the best angles,  ${\rm argmin} \, \{C(\vec{\theta})\, |\,  \vec{\theta} \in \Theta \}$, are found using finite-shot $C(\vec{\theta})$ estimates obtained during the optimization, while $r$ is computed with the infinite-shot $C(\vec{\theta})$ for these angles. 

\begin{figure}[htb]
    \includegraphics[width=\columnwidth]{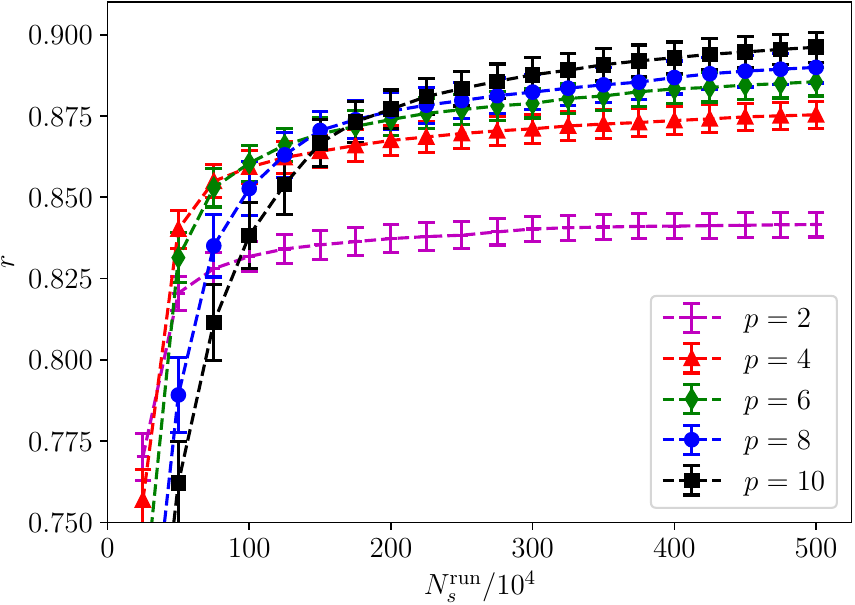}
    \caption{ {\bf The $p$ convergence of the surrogate-based optimization for the 16-qubit 3-regular Max-Cut problems with the infinite-shot cost functions estimates.}  We plot $r$ for $N_s=5000$ optimization runs from Fig.~\ref{fig:MaxCut_pconv}. We  use the infinite-shot cost function estimates that are computed for for the best angles found within the first $N_s^{\rm run}$ run shots. The best angles are found using the finite-shot $C(\vec{\theta})$ estimates, the same as in Fig.~\ref{fig:MaxCut_pconv}. The averaging, and the error bar computation are performed the same as in  Fig.~\ref{fig:MaxCut_pconv}. }
    \label{fig:MaxCut_pconv_shot_inf}
\end{figure}

In Fig.~\ref{fig:MaxCut_shot_inf} we compare the surrogate-based optimization to DARBO, similarly to Fig.~\ref{fig:MaxCut}. Moreover, in Fig.~\ref{fig:MaxCut_pconv_shot_inf} we analyze the results improvement with an increasing QAOA round number, analogically to Fig.~\ref{fig:MaxCut_pconv}. In both cases, we find that the results are qualitatively very similar to the main text results obtained with the finite-shot estimates.

\subsection{QAOA for Random Ising Model on Heavy-Hex Graphs}
\label{sec:heavy_hex_add_res}

\begin{figure}[t]
        \includegraphics[width=0.99\columnwidth]{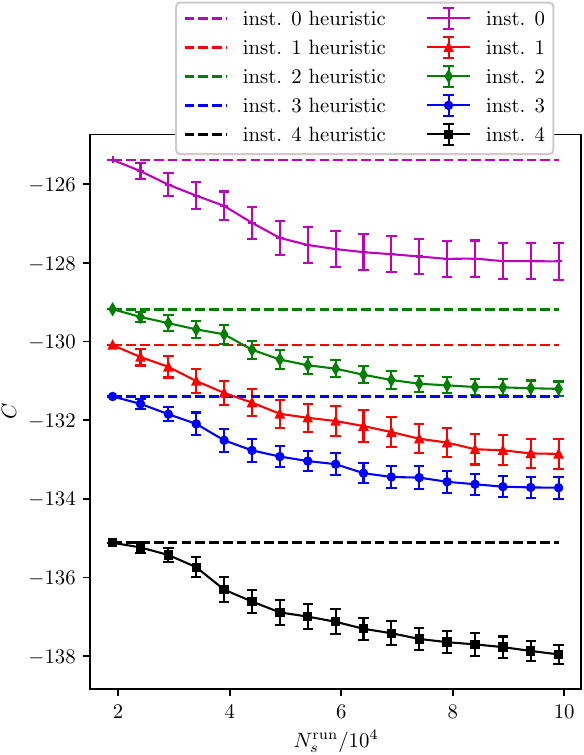}
    \caption{{\bf A convergence of infinite-shot cost function estimates for the MPS optimization of QAOA for heavy-hex 127-qubit random Ising models.}.  Here we plot infinite-shot $C(\vec{\theta})$ estimates for the optimization runs and the heuristic angles, $\vec{\theta}_{\rm heur}^{p=3}$, from Fig.~\ref{fig:hhex_MPS}. We show here results for exactly the same graph instances and runs as in Fig.~\ref{fig:hhex_MPS}. The solid lines are the cost estimates for the best angles found during the first $N_S^{\rm tot}$ shots of an optimization run and the dashed angles are the costs for the heuristic angles. The best optimized angles are found according to the finite-shot, optimized cost estimates,  the same as in Fig.~\ref{fig:hhex_MPS}. For each Hamiltonian instance we average over 32 runs and compute the error bars the same as in Fig.~\ref{fig:hhex_MPS}.   }
    \label{fig:hhex_MPS_shot_inf}
\end{figure}

Analogically to Appendix~\ref{app:MaxCut_add_res}, we use infinite-shot estimates of the cost function to quantify more robustly the improvement of the optimized angles with respect to the heuristic ones. In Fig.~\ref{fig:hhex_MPS_shot_inf}, we plot infinite-shot cost function estimates for the best angles found during the first $N_s^{\rm run}$ shots of the optimization runs from Fig.~\ref{fig:hhex_MPS}, and for the heuristic angles $\theta_{\rm heur}^{p=3}$. The best optimized angles are found according to the finite-shot, optimized estimates of the cost, similarly as in  Appendix~\ref{app:MaxCut_add_res}. We find that the infinite-shot cost function of the optimization runs improves upon the heuristic angles similarly to the results of Fig.~\ref{fig:hhex_MPS}, which are obtained with the finite-shot estimates.

Additionally, in Table~\ref{tab:hhex_MPS_Dconv}, we compare the optimized costs $C_{\rm opt}$ in Eq.~\eqref{eq:Copt} for two different values of the MPS simulations refinement parameter, $D=512$, and $D=1024$, that were used to perform the optimization. The main text results were obtained for $D=1024$. In both cases, the cost function values averaged over the optimization runs agree within $95\%$ confidence intervals estimated with the mean standard deviation. This indicates that the results are likely converged in $D$. We reach the same conclusion for MPS estimates of $C(\vec{\theta}_{\rm heur}^{p=3})$.

\begin{table}[htb]
\begin{tabular}{|c|c|c|c|}
\hline
instance & $D$ & $C(\vec{\theta}_{\rm heur}^{p=3})$ & $C(\vec{\theta}_{\rm opt})$  \\ \hline  
0 & 512 & -125.4(1)  & -129.1(3)  \\ \hline
0 & 1024 & -125.5(2) & -128.8(5) \\ \hline
1 & 512 & -130.0(1) & -133.8(3)  \\ \hline
1 & 1024 &  -130.0(2)  & -133.7(4) \\ \hline
2 & 512 &  -129.3(2) &  -131.9(2)  \\ \hline
2 & 1024 & -129.4(2) &  -132.1(2) \\ \hline
3 & 512 & -131.3(2) &  -134.7(2)  \\ \hline
3 & 1024 & -131.4(1) &  -134.6(3) \\ \hline
4 & 512 & -135.0(1) & -138.6(3)  \\ \hline
4 & 1024 &  -135.1(2) &  -138.8(3) \\ \hline
\end{tabular}
\caption{{\bf A comparison of $D=512$ and $D=1024$ MPS surrogate-based optimization for $p=3$ QAOA and 127-qubit random heavy-hex Ising model.}  Here we compare  the final  optimized cost function $C(\vec{\theta}_{\rm opt})$  averaged over $32$  $D=512$ and $D=1024$ optimization runs for $5$ instances of the Hamiltonian. As a reference, we give $D=512$ and $D=1024$ $C(\vec{\theta}_{\rm heur}^{p=3})$ estimates. The results for $D=1024$ runs are plotted in Fig.~\ref{fig:hhex_MPS}, while the $D=512$ results were obtained in the same way, with the only difference  being $D$ value. For a reference, we also list here $D=512$ and $D=1024$ estimates of $C(\vec{\theta}_{\rm heur}^{p=3})$. Here, $C(\vec{\theta}_{\rm heur}^{p=3})$ is evaluated once per an optimization run,  as $\vec{\theta}_{\rm heur}^3$ is included among runs' initial angles, and is averaged over all its evaluations. We compute the error bars as the mean standard deviation multiplied by a factor of $2$, the same as in Fig.~\ref{fig:hhex_MPS}.  All the cost functions are evaluated with $N_s=1000$ shots. }
\label{tab:hhex_MPS_Dconv}
\end{table}

\section{Bond Dimension Convergence of the MPS Cost Function for the Hardware Runs.}
\label{app:MPS_hard}

\begin{figure*}[t]
    \includegraphics[width=\textwidth]{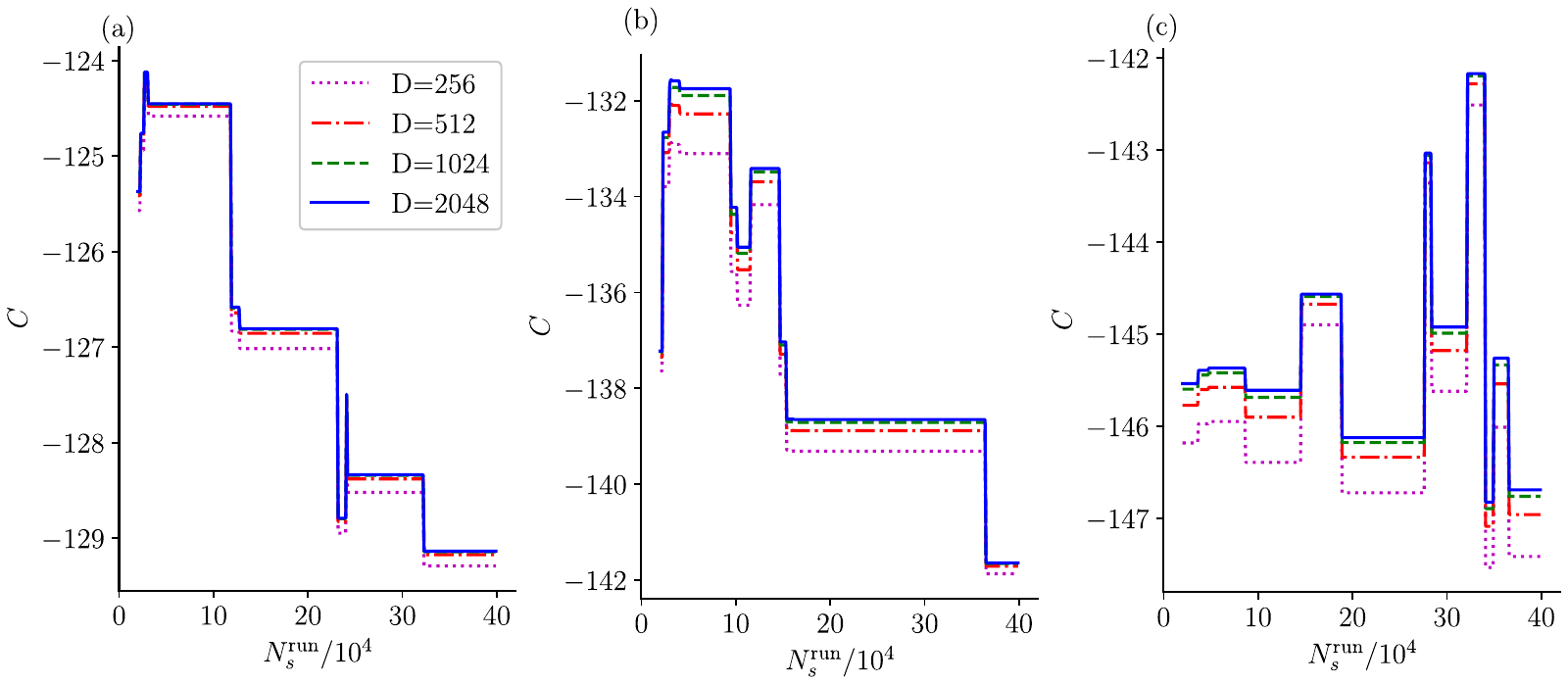}
    \caption{ {\bf Convergence of the MPS cost function with the MPS bond dimension, $D$, for representative \texttt{ibm\_torino} hardware optimization runs.} In this figure, we plot the MPS infinite-shot cost function for hardware runs nb. 0 from Fig.~\ref{fig:torino}, and $p=3$ (a), $p=4$ (b), $p=5$ (c). The $D=2048$ results are plotted in insets of Fig.~\ref{fig:torino}. }
    \label{fig:torino_Dconv}
\end{figure*}

In Fig.~\ref{fig:torino_Dconv}, we plot the MPS cost function for 9 representative \texttt{ibm\_torino} optimization runs from Fig.~\ref{fig:torino}, and for the MPS refinement parameter $D=256,512,1024,2048$. We see that the results converge quickly with an increasing $D$. The observed convergence indicates that $D=2048$, which is used in Fig.~\ref{fig:torino}, is large enough to provide reliable results.

\section{Implementation of Surrogate-based Optimization using \textit{mystic}} \label{app:surr_config}

We use the \textit{mystic} Python software to ``wrap'' an interface for surrogate-based
learning \cite{mckerns2009mystic, mckerns2011building} around a \texttt{cost} function, where we first create a file-based archive to store the training \texttt{data} associated with calling the cost function.

\begin{small}
\begin{verbatim}
# create a data archive, and associate it
# with a cost function
from mystic.cache.archive import file_archive, read
archive = read("truth.db", type=file_archive)
truth = WrapModel("truth", cost, nx=2*p, ny=None,
                  cached=archive)
\end{verbatim}
\end{small}

This creates a callable object, \texttt{truth}, that will log its inputs and output to a file-based \texttt{archive} named \texttt{truth.db}. The keywords \texttt{nx} and \texttt{ny} correspond to the dimensionality of the inputs and output, with \texttt{None} indicating a scalar.

We then sample $N_{\rm init}$ initial datapoints, using uniform random sampling of the cost function, and build a classical \texttt{surrogate} from the training data using RBF interpolation.

\begin{small}
\begin{verbatim}
# generate a dataset by sampling truth
data = truth.sample(bounds, pts=N_init)

# create an inexpensive surrogate for truth
kwds = dict(smooth=0.0, noise=0.0, method="thin_plate")
surrogate = InterpModel("surrogate", nx=2*p, ny=None,
                        data=truth, **kwds)
\end{verbatim}
\end{small}

Here, \texttt{InterpModel} creates a callable object that performs RBF
interpolation on the data associated with \texttt{truth}. A dictionary
of hyperparameters, \texttt{kwds}, is used to configure the interpolation,
and thus the surrogate. We choose a thin-plate basis function, where the
interpolation is constrained to go through the data (i.e. no smoothing,
and no noise). These parameters are not chosen through pre-training, but
instead because they are the most robust settings that enable the surrogate
to reproduce the training data exactly.

We then have an archive of training \texttt{data} of the form $\{ (C(\vec{\theta}), \vec{\theta} ) \, | \, \vec{\theta} \in \Theta \}$, with $\Theta$ being a set of all parameters for which the device cost function $C$ was evaluated, and a surrogate $C^{\rm surr}$ that has not yet been fit to the training data.
The remainder of our approach iteratively fits the surrogate to the training data. Within each iteration, we perform an optimization that solves for the minimum
$\vec{\theta}_{\rm cand}$ of the updated $C^{\rm surr}$, and compares
$C^{\rm surr}(\vec{\theta}_{\rm cand})$ to $C(\vec{\theta}_{\rm cand})$.

Specifically:
\begin{enumerate}
    \item Build an initial set of training data from the quantum computer in the form of $\{ (C(\vec{\theta}), \vec{\theta} ) \}$.
    \item Create an \texttt{InterpModel} to fit to the training data.
    \item Continue to iterate the following, until termination is reached:
    \begin{enumerate}
        \item Fit the surrogate to the training data.
        \item Find the surrogate minimum using Differential Evolution.
        \item Evaluate the cost function at the surrogate minimum.
        \item Update the training data with the new evaluation of cost.
    \end{enumerate}
\end{enumerate}

We perform the minimization on the surrogate with the Differential Evolution
algorithm \texttt{diffev2} from \textit{mystic} \cite{mckerns2009mystic, mckerns2011building}.

\begin{small}
\begin{verbatim}
from mystic.abstract_solver import AbstractSolver
from mystic.termination import EvaluationLimits
loop = AbstractSolver(3) # nx
loop.SetTermination(EvaluationLimits(maxiter=500))
while not loop.Terminated():

    # fit the surrogate to data in truth database
    surrogate.fit(data=data)

    # find the minimum of the surrogate
    results = diffev2(lambda x: surrogate(x), bounds, 
                      bounds=bounds,npop=20*p, gtol=500,
                      ftol=5e-4, full_output=True)

    # evaluate truth at the same input 
    # as the surrogate minimum
    xnew = results[0].tolist()
    ynew = truth(xnew)
\end{verbatim}
\end{small}

We configure \texttt{diffev2} to have \texttt{bounds} as described in Sec.~\ref{sec:dev_opt}, a population \texttt{npop} of $20p$ for $p$ QAOA layers, and a termination when the cost function does not change more than \texttt{ftol} over \texttt{gtol} iterations. We note that we rescale the cost function by the number of graph vertices $n$, to ensure approximately constant scaling of the allowed cost function values with $n$.  
We set this termination condition for our real device optimizations to not
exceed $N_{\text{it}}$ iterations (including the initial sampling of points).

\clearpage

\end{document}